\documentclass[usenatbib, twocolumn, a4paper,nofootinbib]{aastex631}
\usepackage[utf8]{inputenc}  
\usepackage[T1]{fontenc}
\usepackage{graphicx,psfrag}
\usepackage{mathrsfs}
\usepackage{amsmath,amsfonts,amssymb,pifont,gensymb}
\usepackage{multirow,enumerate}
\usepackage{xcolor}
\usepackage{acronym}
\usepackage{xspace}
\usepackage[normalem]{ulem}
\usepackage{mathtools}
\usepackage{makecell}
\usepackage{tabularray}

\newcommand{\dett}{\textrm{det}}
\newcommand{\dd}{\mathrm{d}}
\newcommand{\HL}{H_L}
\newcommand{\HNL}{H_{U}}
\newcommand{\HU}{\HNL}

\newcommand{\data}{\vec{d}}

\newcommand{\Nset}{N_{\rm set}}
\newcommand{\Nimg}{N_{\rm img}}

\newcommand{\PLUdet}{\mathcal{P}^L_{U}}
\newcommand{\PLU}{P^L_{U}}

\newcommand{\RLU}{R^L_{U}}

\newcommand{\TLU}{T^L_{U}}

\newcommand{\BLUdet}{\mathcal{B}^L_{U}}
\newcommand{\BLU}{B^L_{U}}
\newcommand{\OLU}{O^L_{U}}

\newcommand{\Rldet}{\mathcal{R}_L}
\newcommand{\Rudet}{\mathcal{R}_U}
\newcommand{\Ndet}{\mathcal N}

\newcommand{\Rl}{R_L}
\newcommand{\Ru}{R_U}
\newcommand{\Rtot}{R}
\newcommand{\N}{N}
\newcommand{\Nl}{N_L}
\newcommand{\Nu}{N_U}
\newcommand{\Navg}{\langle N \rangle}
\newcommand{\Nlavg}{\langle N_L \rangle}
\newcommand{\Nuavg}{\langle N_U \rangle}

\newcommand{\vtheta}{\vec \theta}
\newcommand{\vthetal}{\vec \theta_L}
\newcommand{\vy}{\vec y}
\newcommand{\vnui}{\vec \nu_i}

\begin{document}
\title{Strong gravitational-wave lensing posterior odds} 

\author[0000-0002-3887-7137]{Otto A. Hannuksela}
\affiliation{Department of Physics, The Chinese University of Hong Kong, Shatin, New Territories, Hong Kong}
\email{oahannuksela@cuhk.edu.hk}
\author[0000-0001-5597-8098]{K. Haris}
\affiliation{Department of Physics, National Institute of Technology, Kozhikode, Kerala 673601, India}
\email{harismk@nitc.ac.in}
\author[0000-0002-0471-3724]{Justin Janquart}
\affiliation{Centre for Cosmology, Particle Physics and Phenomenology - CP3,
Universite Catholique de Louvain, Louvain-La-Neuve, B-1348, Belgium}
\affiliation{Royal Observatory of Belgium, Avenue Circulaire, 3, 1180 Uccle, Belgium}
\author[0000-0002-0471-3724]{Harsh Narola}
\affiliation{Nikhef – National Institute for Subatomic Physics, Science Park 105, 1098 XG Amsterdam, The Netherlands}
\affiliation{Institute for Gravitational and Subatomic Physics (GRASP), Utrecht University, Princetonplein 1, 3584 CC Utrecht, The Netherlands}
\author[0000-0002-0471-3724]{Hemantakumar Phurailatpam}
\affiliation{Department of Physics, The Chinese University of Hong Kong, Shatin, New Territories, Hong Kong}
\author[0000-0003-3600-2406]{Jolien D. E. Creighton}
\affiliation{Leonard E. Parker Center for Gravitation, Cosmology, and Astrophysics, University of Wisconsin-Milwaukee, Milwaukee, WI 53201, USA}
\author[0000-0003-3600-2406]{Chris Van Den Broeck}
\affiliation{Nikhef – National Institute for Subatomic Physics, Science Park 105, 1098 XG Amsterdam, The Netherlands}
\affiliation{Institute for Gravitational and Subatomic Physics (GRASP), Utrecht University, Princetonplein 1, 3584 CC Utrecht, The Netherlands}

\begin{abstract}
\noindent
Like light, gravitational waves are gravitationally lensed by intervening massive astrophysical objects, such as galaxies, clusters, black holes, and stars, resulting in a variety of potentially observable gravitational-wave lensing signatures. 
Searches for gravitational-wave lensing by the LIGO-Virgo-KAGRA (LVK) collaboration have begun. 
One common method focuses on strong gravitational-wave lensing, which produces multiple \textquotedblleft images\textquotedblright: repeated copies of the same gravitational wave that differ only in amplitude, arrival time, and overall \textquotedblleft Morse phase.\textquotedblright   
The literature identifies two separate approaches to identifying such repeated gravitational-wave events based on frequentist and Bayesian approaches. 
Several works have discussed selection effects and identified challenges similar to the well-known \textquotedblleft birthday problem\textquotedblright, namely, the rapidly increasing likelihood of false alarms in an ever-growing catalogue of event pairs. 
Here, we discuss these problems, unify the different approaches in Bayesian language, and derive the posterior odds for strong lensing. 
In particular, the Bayes factor and prior odds are sensitive to the number of gravitational-wave events in the data, but the posterior odds are insensitive to it once strong lensing time delays are accounted for. 
We confirm the Lo et al. (2023) finding that selection effects enter the Bayes factor as an overall normalisation constant. 
However, this factor cancels out in the posterior odds and does not affect frequentist approaches to strong lensing detection. 
\end{abstract}

\section{Introduction}

Gravitational waves, like light, are predicted by general relativity to be gravitationally lensed by massive intervening objects such as galaxies, clusters, stars, and black holes~\citep{Takahashi2003WaveBinaries}. 
In the strong lensing regime—typically involving massive galaxies or clusters—a gravitational wave can be split into multiple \textquotedblleft images,\textquotedblright i.e., repeated copies of the same event, arriving at different times~\citep{Haris2018IdentifyingMergers}. 
Less massive lenses (e.g., stars~\citep{Diego2019ConstrainingFrequencies,Cheung2020Stellar-massWaves,Mishra2021GravitationalGalaxies,Yeung2021MicrolensingMacroimages,Shan2022WaveStudy,Shan2023MicrolensingWaves}, intermediate-mass black holes~\citep{Lai2018DiscoveringLensing,Gais2022InferringStatistics,Wright2021Gravelamps:Selection}, or compact halo objects~\citep{Basak2022ConstraintsMicrolensing,Liu2023ExploringConfigurations}) can produce millisecond-separated signals that interfere and/or diffract, leading to frequency-dependent waveform distortions~\citep{cao2014gravitational,Lai2018DiscoveringLensing}. 
Since the first gravitational-wave detection in 2015~\citep{LIGOScientific:2016aoc} and subsequent detector improvements~\citep{LIGOScientific:2014pky,VIRGO:2014yos,KAGRA:2020tym}, theoretical studies are now predicting strong lensing detections at a rate of roughly 1 in 1000 events~\citep{Ng2017PreciseHoles,Oguri2018EffectMergers,Wierda2021BeyondLensing,Xu2021PleasePopulations,More2021ImprovedEvents,Smith2022DiscoveringObservatory,Phurailatpam2024LerSimulator}, within the observable reach of current detectors as the size of the current gravitational-wave catalogue keeps increasing~\cite{KAGRA:2021vkt}. 
While no conclusive lensed events have been observed, these non-detections constrain lens populations and high-redshift merger rates~\citep{Hannuksela2019SearchEvents,McIsaac2019SearchRuns,Dai2020SearchO2,Liu2020IdentifyingVirgo,Pang2020LensedDetection,Abbott2020GW190425:M,TheLIGOScientificCollaboration2020GW190521:M_odot,Kim2022Deep-2,Janquart2023Follow-upSearches}. 
Detecting lensed gravitational waves would enable a range of astrophysical and cosmological applications, including measuring galaxy velocity dispersions~\citep{Xu2021PleasePopulations}, probing dark matter~\citep{Basak2022ConstraintsMicrolensing,Jana:2024dhc,barsode2024constraints}, detecting intermediate-mass black holes~\citep{Lai2018DiscoveringLensing}, constraining binary formation channels~\citep{Leong2024ConstrainingWaves}, studying binary environments~\citep{Ubach:2025dob}, testing general relativity~\citep{Goyal2020TestingSignals,Finke2021ProbingBinaries,Hernandez2022MeasuringMergers,Iacovelli2022ModifiedFunction,Narola:2023viz}, investigating galaxy lenses~\citep{Poon2024GalaxyDegeneracy}, localising binaries to sub-galactic precision~\citep{Hannuksela2020LocalizingLensing,Wempe2022AObservations,Shan2023MicrolensingWaves,Uronen2024FindingMulti-messenger}, constraining the cosmological expansion~\cite{Jana2022CosmographyHoles,Jana:2024uta} and enabling multimessenger studies~\citep{Smith:2025axx}.

\begin{figure}
    \centering
    \includegraphics[width=\columnwidth]{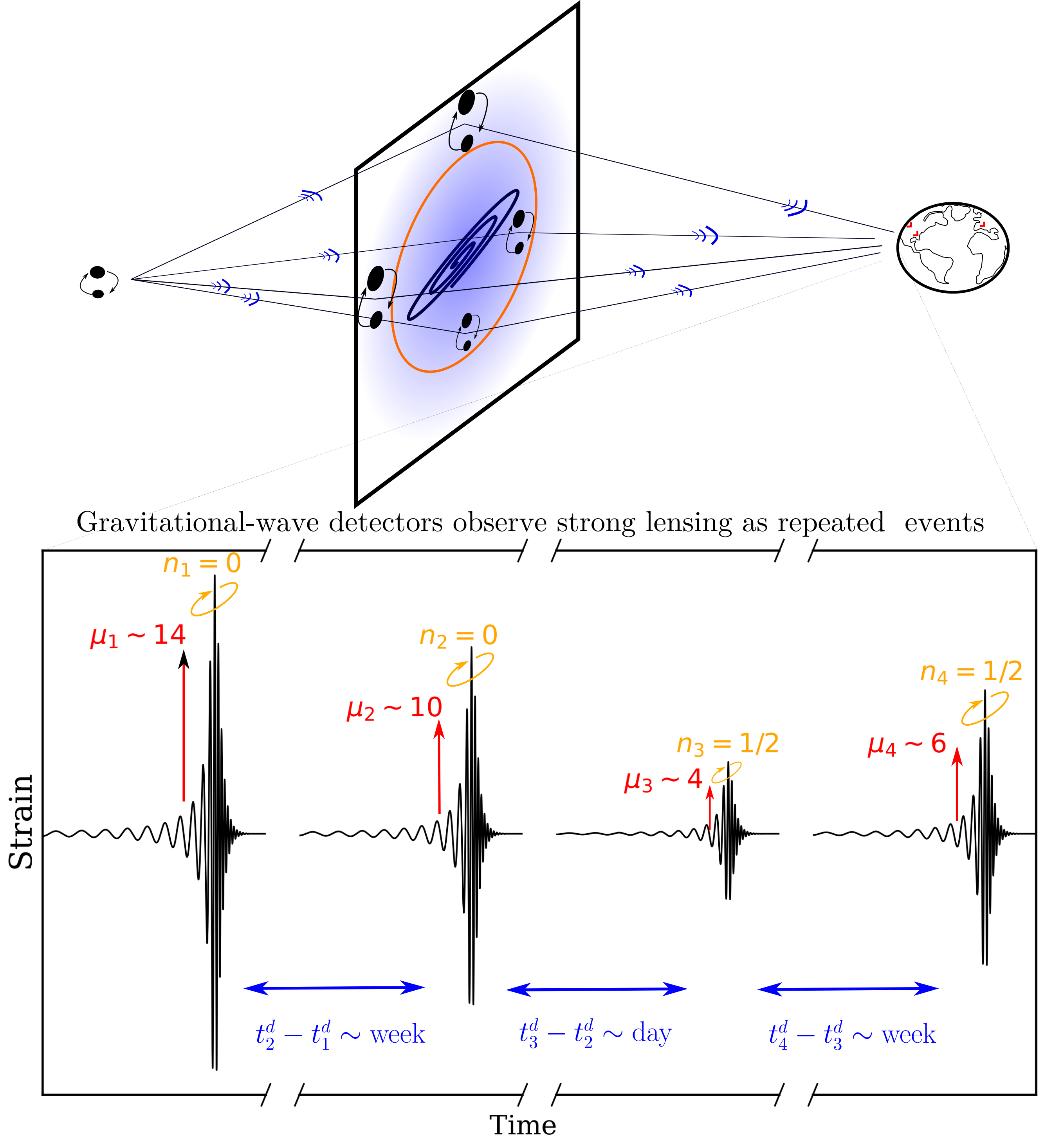}
    \caption{Strong lensing produces multiple images (top panel), which can be detected with the ground-based gravitational-wave detectors as repeated events arriving at different times (bottom panel). 
    The repeated gravitational waves ($i=1,2,3,\cdots$) are otherwise identical, 
    but strong lensing can magnify the gravitational-wave amplitude by a factor $|\mu_i|^{1/2}$ (red), 
    delay their arrival time by a time $t_i^d$ (blue), 
    and induce an overall complex phase shift (orange), referred to as the Morse phase $\pi n_i$. 
    The outcome is $\Nimg$ gravitational waves with different amplitudes and (possibly) overall phases, arriving at the gravitational-wave detectors at different times. 
    Strong lensing searches attempt to identify these repeated events. 
    }
    \label{fig:strong_lensing_illustration}
\end{figure}

While gravitational lensing produces various signatures, this work focuses on strong lensing, which generates multiple images of the gravitational-wave source. 
The primary known strong lensing targets, based on the estimate of strong lensing rates for stellar merger remnants, are galaxies and clusters~\citep{Smith2017WhatClusters,Ng2017PreciseHoles,Li2018GravitationalPerspective,Smith2018DeepGW170814,Smith2018Strong-lensingClusters,Smith2022DiscoveringObservatory,Wierda2021BeyondLensing,Xu2021PleasePopulations,More2021ImprovedEvents,Meena2022GravitationalPopulation,Phurailatpam2024LerSimulator}, which produce these images. 
In electromagnetic astronomy, such images have been directly observed for decades (e.g., lensed quasars)~\citep{Schneider1992GravitationalLenses,Dodelson2017GravitationalLensing,Congdon2018PrinciplesLensing}. For gravitational waves, current detectors lack the angular resolution to resolve images; instead, lensed images appear as repeated, temporally separated by minutes to months (or even years) but otherwise nearly identical signals (Fig.~\ref{fig:strong_lensing_illustration}, bottom panel).

Strong lensing detection strategies include both frequentist and Bayesian approaches. 
The first Bayesian derivation used posterior distributions to compute the Bayes factor, assuming repeated images share detector-frame parameters~\citep{Haris2018IdentifyingMergers}. 
This work used the Bayes factor as a ranking statistic, instead of interpreting it literally, to quantify a frequentist-style p-value test, which was also used by~\cite{Dai2020SearchO2}, who included the so-called Morse phase information. 
In a later work, the Bayes factor was interpreted in a more traditional manner using joint parameter estimation of signal pairs, which improved the accuracy of the Bayes factor computation~\citep{Liu2020IdentifyingVirgo} and included selection effects through the so-called Malmquist prior.\footnote{Malmquist prior was also used in~\cite{Barsode:2024zwv}.} 
This method was extended with explicit lens and binary black hole population models, and it was argued that selection effects should enter as an overall normalisation factor, not through the Malmquist prior, seemingly disagreeing with earlier work~\citep{Lo2021ASignals}. 
Later work showed that these Bayes factors are highly sensitive to the population model assumptions, that the parametric population models can result in misleading results, and advocated for non-parametric instead of parametric models~\citep{Cheung2023MitigatingMethods}. 
Other works highlighted that false alarms in strong lensing searches grow rapidly~\citep{Calskan2022LensingWaves}, a problem mitigated by incorporating lensing time delays to weed out lensing candidate pairs with long time delays~\citep{Wierda2021BeyondLensing}. 
These false alarm issues were later incorporated in the frequentist-style p-value estimates~\citep{Barsode:2024zwv}. 
However, this false alarm growth was not evident in the Bayesian formulation. Contrary to these previous works,~\cite{Diego2021EvidenceLIGO-Virgo} argued that Bayes factors without selection effects or population models provide a more robust detection metric. 
Thus, the roles of selection effects, population priors, frequentist versus Bayesian approaches, and rising false alarms have been a subject of significant discussion.

Here we argue that the posterior odds is the definitive criterion for strong lensing detection, as it remains independent of selection effects and catalogue size for a fixed population model. 
Although selection effects enter the Bayes factor as an overall normalisation, they are exactly cancelled by corresponding terms in the prior odds, leaving the posterior odds unaffected (see Sections~\ref{sec:bayes_factor} and~\ref{sec:odds_ratio}). 
While the prior odds in favour of lensing decrease as more events are observed, this effect is compensated by the increasing ratio of arrival time priors in the Bayes factor, resulting in a stable posterior odds (see Sections~\ref{sec:prior_odds} and~\ref{sec:odds_ratio}). 
Accurate detection statements require explicit modelling of the intrinsic binary black hole and lens populations, as well as the lensing time-delay distribution; neglecting these leads to unreliable inference (see Sections~\ref{sec:prior_odds} and~\ref{sec:odds_ratio}). 
Finally, frequentist methods based on ranking statistics or p-values are not impacted by these Bayesian considerations, as the relevant normalisations cancel out (see Section~\ref{sec:bayes_factor}). 
These results clarify the correct Bayesian framework for strong lensing searches and resolve previous ambiguities regarding the roles of selection effects, population priors, and catalogue size in Bayesian inference (see Section~\ref{sec:discussion}).

\begin{table*}
    \centering
	\begin{tblr}{colspec={l Q[wd=14.5cm] l}} 
        \hline
        \hline
        \textbf{Notation} &  \textbf{Description} & \textbf{Eq.} \\
        \hline
        \hline
         $\HL$ & Set of $\Nset=\Nimg$ signals that are strongly lensed (Sec.~\ref{ssec:strong_lensing_hyp}) & \eqref{eq:hl_definition} \\ 
         $\HU$ & Set of $\Nset=\Nimg$ signals that are not strongly lensed (Sec.~\ref{ssec:strong_lensing_hyp}) & \eqref{eq:hu_definition} \\
         $\dett$ & Denotes that a set of $\Nset=\Nimg$ signals is detectable (Sec.~\ref{ssec:evidence_lensed}) & \eqref{eq:joint_evidence_conditioned} \\ 
         \hline
         $\BLUdet$ & Strong lensing Bayes factor (Sec.~\ref{sec:bayes_factor}) & \eqref{eq:blu_definition} \\
         $\BLU$ & Strong lensing Bayes factor without selection effects (Sec.~\ref{sec:bayes_factor}) & \eqref{eq:bludet_definition} \\
         $\PLUdet$ & Strong lensing prior odds (Sec.~\ref{sec:prior_odds}) & \eqref{eq:prioroddsdet} \\
         $\PLU$ & Strong lensing prior odds without selection effects (Sec.~\ref{sec:prior_odds}) & \eqref{eq:priorodds} \\
         $\RLU$ & Ratio of arrival time priors without selection effects (Sec.~\ref{sec:odds_ratio}) & \eqref{eq:rlu_defnition} \\ 
         $\widetilde{B}^L_U$ & Time-independent Bayes factor ($\BLU=\widetilde{B}^L_U\times \RLU$) (Sec.~\ref{sec:odds_ratio}) & \eqref{eq:blu_tilde_definition} \\
         $\TLU$ & Rate odds ($\TLU= \RLU \times \PLU$) (Sec.~\ref{sec:odds_ratio}) & \eqref{eq:tlu_definition} \\
         \hline
         $\mathcal \Rl$ & Rate of lensed, detectable mergers & \eqref{eq:incorrect_prior_odds} \\
         $\Rl$ & Intrinsic rate of lensed mergers (detected \emph{and} not detected mergers) & \eqref{eq:rl_definition} \\
         $\mathcal \Ru$ & Rate of non-lensed, detectable mergers & \eqref{eq:incorrect_prior_odds} \\ 
         $\Ru$ & Intrinsic rate of non-lensed mergers (detected \emph{and} not detected mergers) & \eqref{eq:ru_definition} \\ 
         $\mathcal \Rtot$ & Rate of total, detectable mergers & \eqref{eq:incorrect_prior_odds} \\ 
         $\Rtot$ & Intrinsic rate of total mergers (detected \emph{and} not detected mergers) & \eqref{eq:Rtot_definition} \\ 
         $T$ & Observation time  & (\ref{eq:nl_nu_avg_definition}) \\
         $\Nlavg$ & $\Nlavg=\Rl T$ is the intrinsic, expected number of lensed mergers (detected \emph{and} not detected mergers) & \eqref{eq:rl_definition}\\
         $\Nuavg$ & $\Nuavg=\Ru T$ is the intrinsic, expected number of non-lensed mergers (detected \emph{and} not detected mergers) & \eqref{eq:ru_definition}\\ 
         $\Nlavg \Nimg$ & Intrinsic, expected number of strongly lensed signals & \eqref{eq:number_lensed_signals_not_mergers} \\
         $\Nimg$ & Number of strong lensing images ($\Nimg=\Nset$) & \eqref{eq:gw_lensed} \\ 
         \hline
         $\vtheta$ & $\vtheta := \{m_1,m_2,\vec{s}_1,\vec{s}_2, z_s, \theta_{\mathrm{JN}}, \alpha, \delta, \psi, \phi_c, t_u\}$ denotes the set of binary parameters & \eqref{eq:gw_lensed}\\ 
         $\vthetal$ & Denotes the lens parameters (lens redshift, Einstein radius, ellipticity, ...)  & \eqref{eq:gw_lensed}\\ 
         $\vy$ & Binary source position with respect to the lens line-of-sight to the lens & \eqref{eq:gw_lensed} \\ 
         $\vnui$ & $\vnui(\vthetal, \vy, z_l, z_s):=\{\mu_i, t_i^d, n_i\}$ denotes the image properties of the $i^{\mathrm{th}}$ strong lensing image, with $\mu_i$, $t_i^d$, and $n_i$ being the magnification, arrival time, and Morse phase of the $i^\text{th}$ image & \eqref{eq:gw_lensed} \\ 
         $t_i$ & Denotes the arrival times of the $i^\text{th}$ signal ($t_i=t_u$ under the unlensed hypothesis and $t_i=t_i^d+t_u$ under the lensed hypothesis) & \eqref{eq:gw_lensed} \\ 
         \hline
         \hline
    \end{tblr}
    \caption{Summary of the nomenclature used in this work. }
    \label{tab:notations}
\end{table*}

\section{Strong lensing hypothesis: Repeated events} \label{sec:strong_lensing_hypothesis}
Here, we will review the basic ideas and formalism behind the strong lensing hypothesis. 
For a comprehensive derivation of the effect of lensing on gravitational waves, we refer the reader to~\citet{Takahashi2003WaveBinaries}. 
The first Bayesian derivation of the strong lensing hypothesis was performed by~\citet{Haris2018IdentifyingMergers}, 
which was later refined in a hierarchical analysis by~\citet{Lo2021ASignals}. 
Besides these works, many other studies have investigated strong lensing detection in various forms~\citep{Hannuksela2019SearchEvents,Li2019FindingEvents,McIsaac2019SearchRuns,Hannuksela2020LocalizingLensing,Dai2020SearchO2,Liu2020IdentifyingVirgo,Janquart2021AEvents,Janquart2021OnModes,TheLIGOScientificCollaboration2021SearchRun,LIGOScientific:2023bwz,Janquart2024WhatHypothesis,Chakraborty:2024net,Chakraborty:2025maj}.

\subsection{Image formation and Morse theory}

The basic picture is that a fraction of binary black hole mergers will be lensed by massive astrophysical objects. 
Strong lensing occurs if the merger aligns closely enough with the lens's line-of-sight. 
A typical example object that produces strong lensing is either a galaxy or a galaxy cluster. 
In this case, the strong lensing produces multiple \textquotedblleft images\textquotedblright of the binary black hole merger (Fig.~\ref{fig:strong_lensing_illustration}, top panel). 

In the electromagnetic band, such multiple images can be visible, having resolvably different positions on the sky, with one good example being strongly lensed quasars, which scientists have observed for decades~\citep{Schneider1992GravitationalLenses,Dodelson2017GravitationalLensing,Congdon2018PrinciplesLensing}. 
However, the picture is different for gravitational waves.
In particular, the current gravitational-wave detectors cannot discern the images in the sky due to the super-degree sky resolution ($\mathcal O(1-10\,\text{deg}^2)$ for signals confidently detected in $\geq$3 detector networks)~\cite[e.g.][]{2009NJPh...11l3006F}. 
Instead, these gravitational-wave images could be observed as \textquotedblleft repeated events\textquotedblright separated in time by minutes to months (years) when lensed by galaxies (galaxy clusters) in the ground-based gravitational-wave detectors (Fig.~\ref{fig:strong_lensing_illustration}, bottom panel)~\cite[e.g.][]{Phurailatpam2024LerSimulator}.

\subsection{Gravitational-wave observables}
Because strong lensing does not alter the frequency evolution of the gravitational waves,\footnote{This only applies when the Schwarzschild radius of the lens is much larger than the wavelength of the gravitational wave, referred to as the geometrical optics limit~\citep{Takahashi2003WaveBinaries}.}
the repeated events are identical otherwise but may differ in amplitude, phase, and arrival time. 
In other words, strong lensing can only magnify, delay, or phase-shift the original gravitational wave~\citep{Dai2017OnWaves,Ezquiaga2022ModifiedLensing}. 

\begin{figure*}
    \centering
    \includegraphics[width=\textwidth]{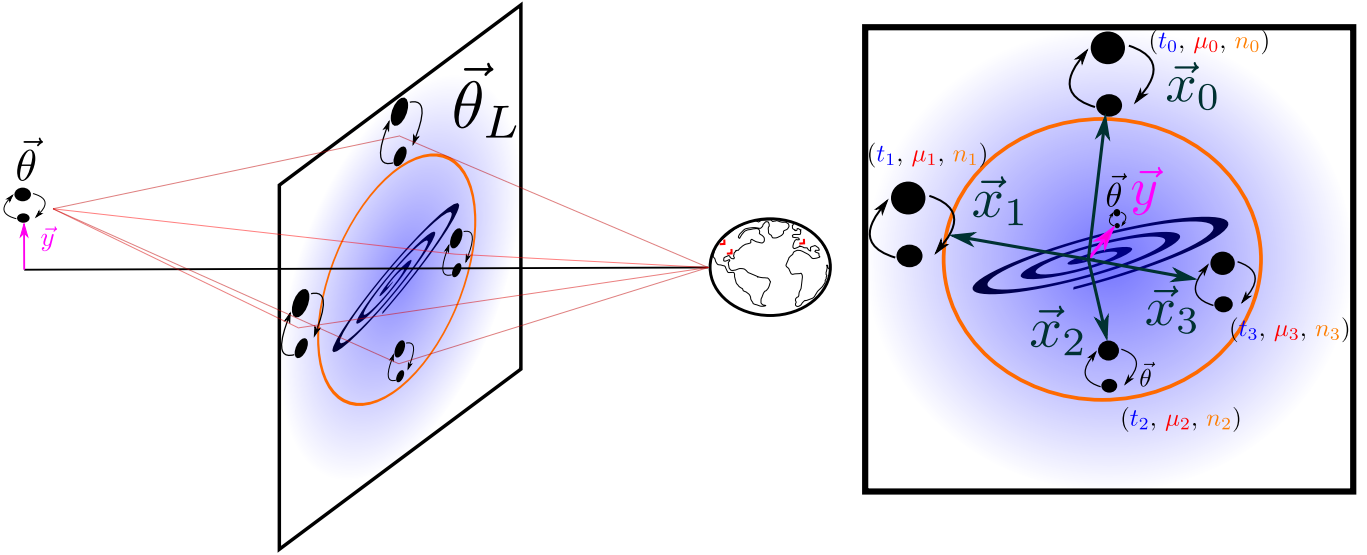}
    \caption{A binary black hole is strongly lensed by a galaxy, producing multiple images. 
    The galaxy lens mass distribution is parametrised with $\vthetal$, the binary black hole orbital parameters and position with $\vtheta$ and $\vy$, respectively. 
    These multiple images each have a position $\vec x_i$, magnification $\mu_i$, arrival time $t_i$, and Morse phase $n_i$ associated with them. 
    These parameters are collectively used to describe the lensed gravitational waves. 
    }
    \label{fig:strong_lensing_configuration}
\end{figure*}
More specifically, strong lensing will produce $\Nimg$ binary black hole images, depending on the lens configuration and the position of the binary with respect to the lens (typically 2 or 4 images for a strong lensing galaxy). 
The $i^\mathrm{th}$ gravitational-wave image will arrive with some time delay due to lensing $t_i^d$, with some magnification $\mu_i$, and some overall Morse phase shift $\pi n_i$, satisfying, in the Fourier space~\citep{Takahashi2003WaveBinaries}~\footnote{Note that we assume the engineering convention for Fourier transforms.}
\begin{equation}\label{eq:gw_lensed}
\begin{split}
    h_{L,+,\times}^{i}(f;\vtheta,\vthetal,\vy,z_l,z_s) = &|\mu_i|^{1/2} e^{i (2 \pi f t_i^d - \pi n_i \text{sign}(f))} \\
    &\times h_{+,\times}(f;\vtheta)\,,
\end{split}
\end{equation}
where $h_{+,\times}(f;\vtheta)$ is the plus/cross polarization of the gravitational-wave in the absence of strong gravitational-wave lensing, $\vtheta$\footnote{$\vtheta := \{m_1,m_2,\vec{s}_1,\vec{s}_2, z_s, \theta_{\mathrm{JN}}, \alpha, \delta, \psi, \phi_c, t_u\}$,\footnote{$m_{1,2}$} denote the component masses, $\vec{s}_{1,2}$ the black hole spins, $z_s$ the source redshift, $\theta_{\mathrm{JN}}$ the inclination between total angular momentum $\vec J$ and the direction of propagation $\vec N$, $\alpha$ and $\delta$ are the right ascension and declination, $\psi$ the polarization angle of the GW, $\phi_{c}$ the phase of coalescence,  and $t_u$ the arrival time of the unlensed GW at the detector} are the binary black hole source parameters. 
The Morse index $n_i$ equals $0$, $1/2$, and $1$ for the minima, saddle, and maxima points of the lensing time delays and can be solved when the lens geometry is known. 

The only lens-related observables in the gravitational-wave detector are the image properties of each image $\vnui:=\{t_i^d, \mu_i, n_i \}$ [cf. Eq.~\ref{eq:gw_lensed}]. 
These image properties depend on the lens parameters $\vthetal$, the source position $\vy$, the lens redshift $z_l$, and the source redshift $z_s$. 
Indeed, the image properties can be solved using standard tools in the lensing literature, given the lens properties. 
Mathematically, we can express the mapping  as  
$\vnui \rightarrow \vnui(\vthetal, \vy, z_l, z_s)$. 

The precise mapping depends on the lens configuration and therefore the lens model.\footnote{We refer the reader to standard lensing textbooks such as~\citep{Schneider1992GravitationalLenses,Dodelson2017GravitationalLensing,Congdon2018PrinciplesLensing}, for a review of various lens models that can be used.}
Therefore, once we choose a lens model with some parameters $\vthetal$, we can express the lensed gravitational-wave from the $i^{\mathrm{th}}$ image as:
\begin{equation}
    h_{L,+,\times}^{i}(f;\vtheta, \vthetal, \vy, z_l, z_s) = F_i(f;\vnui(\vthetal, \vy, z_l, z_s) ) h_{+,\times}(f;\vtheta)\,,
\end{equation}
where 
\begin{equation}
F_i(f;\vnui(\vthetal, \vy, z_l, z_s))=|\mu_i|^{1/2} e^{i 2 \pi f t_i^d - i \pi n_i} 
\end{equation}
is the amplification factor that now encodes the information about the lens configuration through the image properties $\vnui(\vthetal,\vy,z_l, z_s)$, which can be related to the lens configuration by solving the lens equation~\citep[e.g.][]{Schneider1992GravitationalLenses,Dodelson2017GravitationalLensing,Congdon2018PrinciplesLensing}. 

For a given gravitational-wave detector and image $i$, the gravitational-wave strain 
\begin{equation}
\begin{split}
	&h_L^i(f;\vtheta, \vthetal, \vy, z_l, z_s)\\
	&=\mathrm F_+(\alpha,\delta,\psi,t_i) h_{L,+}^i(f;\vtheta, \vthetal, \vy, z_l, z_s)\\
    &+ \mathrm F_\times(\alpha,\delta,\psi,t_i)h_{L,\times}^i(f;\vtheta, \vthetal, \vy, z_l, z_s)\,,
\end{split}
\end{equation}
where $\mathrm F_+$ and $\mathrm F_\times$ are the standard beam pattern functions that encode the detector's response to the passing gravitational wave that depend on the GW sky location (right-ascension $\alpha$ and declination $\delta$), the GW polarization $\psi$, and, due to Earth's rotation, the arrival time at the detector $t_i=t_u+t_i^d$. 
These are available in standard gravitational wave analysis suites (e.g., \texttt{LALSuite}~\cite{lalsuite}). 

According to Morse theory, the number of minima $N_I$, saddle $N_{II}$, and maxima $N_{III}$ images  satisfy~\citep{Schneider1992GravitationalLenses,Dodelson2017GravitationalLensing,Congdon2018PrinciplesLensing}
\begin{equation}
    N_I-N_{II}+N_{III}=1\,,
\end{equation}
for non-singular strong lensing configurations, although typically the type-III image is heavily demagnified and therefore not detectable except in some rare strong lensing configurations where, for example, the lens galaxy hosts a shallow core density~\citep{Collett2015CompoundImages,Collett2016TestingLensing,Collett2017Core1}. 
This Morse theory is important because it sets the image types and, therefore, the Morse phases expected in gravitational-wave strong lensing detections, independently of the lens model. 

The notation is summarised in Table~\ref{tab:notations} (see also Fig.~\ref{fig:strong_lensing_configuration}, for illustration).
Throughout this work, we adopt the Planck18 cosmology~\citep{Planck:2018nkj}.

\subsection{Strong lensing hypothesis} \label{ssec:strong_lensing_hyp}

Given that the $i^{\mathrm{th}}$ image arrives in the gravitational-wave detector within some stretch of data $d_i$, the strong lensing hypothesis
\begin{equation} \label{eq:hl_definition}
    \HL : d_i(f) =\text{n}_i(f) + h_L^i(f;\vtheta, \vthetal, \vy, z_l, z_s)\,,
\end{equation}
where $\text{n}_i(f)$ denotes noise (not to be confused with the morse phase $n_i$ of the $i^\text{th}$ image), $h_L^i$ is the $i^{\mathrm{th}}$ gravitational-wave image of a strongly lensed binary black hole merger with binary parameters $\vtheta$, lens parameters $\vthetal$ and source position $\vy$. 
The unlensed hypothesis, on the other hand, states that each dataset $d_i$ contains an unrelated gravitational-wave signal, i.e., 
\begin{equation} \label{eq:hu_definition}
    \HU : d_i(f) =\text{n}_i(f) + h(f;\vtheta_i)\,,
\end{equation}
where $h(f;\vtheta_i)$ is an unlensed gravitational-wave signal from a different binary black hole such that $\vtheta_i\neq \vtheta_j$ (for $i\neq j$). 
One can perform Bayesian model selection with the above hypotheses in principle. 

The Bayesian model selection method uses hierarchical inference~\citep{Lo2021ASignals}, which performs the parameter estimation with the image parameters instead of the lens parameters and then hierarchically reconstructs the lens parameters. 
Here, we assume that the hierarchical inference has already been performed such that we can work directly with the lens parameters $\vthetal$ and $z_l$, instead of the image parameters $\vnui$ or the so-called effective parameters. 
Nevertheless, we encourage readers interested in the full joint inference to read~\citet{Lo2021ASignals}, which considerably expands upon the hierarchical inference aspect. 
The above assumptions and the existing tools have established everything necessary to perform Bayesian analysis with strongly lensed gravitational waves. 

\section{Bayes factor} \label{sec:bayes_factor}

The Bayes factor is a common tool in Bayesian model selection, allowing us to compare the likelihood of two competing hypotheses given the data. 
Several recent works have discussed the definition of the strong lensing Bayes factor, particularly how selection effects and the black hole/lens population contribute to it. 
Early work used the Bayes factor as a ranking statistic~\citep{Haris2018IdentifyingMergers,Hannuksela2019SearchEvents,Dai2020SearchO2}. 
In a later analysis, the Bayes factor was interpreted in a more traditional manner using so-called joint parameter estimation, which improved the accuracy of the Bayes factor computation~\citep{Liu2020IdentifyingVirgo}, including selection effects through the so-called Malmquist prior.

On the other hand, the effect of the binary black hole and lens model on the Bayes factor was studied by~\citet{Lo2021ASignals}. 
In particular,~\citet{Lo2021ASignals} argued that the Bayes factor as defined in prior work needed to be normalised with an overall term accounting for selection effects in a different manner from the approach using the Malmquist prior. 
The work gave rise to two different terminologies: 
The coherence ratio (Bayes factor without selection effects as it is defined in~\citet{Haris2018IdentifyingMergers,Hannuksela2019SearchEvents,Dai2020SearchO2,Liu2020IdentifyingVirgo}) 
and the Bayes factor (the coherence ratio multiplied by the normalisation constant as defined in~\citet{Lo2021ASignals}). 
The coherence ratio-Bayes factor terminology was also used in the first LIGO--Virgo collaboration search for gravitational-wave lensing~\citep{TheLIGOScientificCollaboration2021SearchRun}. 

However, the role of selection effects and population prior in the Bayes factor has been debated. 
In particular, the approach employing the Malmquist prior~\citep{Liu2020IdentifyingVirgo} seemingly disagrees with the alternative interpretation of selection effects as an overall normalisation to the Bayes factor. 
Furthermore, the importance of the population information on the lensed detections has also been a subject of discussion; 
for example, later work argued that the coherence ratio using an uninformative mass prior and without selection effects was a better discriminator of whether an event is lensed or not~\citep{Diego2021EvidenceLIGO-Virgo}. 

In this section, we contrast the various definitions of the Bayes factor and give our derivation and interpretation. 
Our findings largely agree with~\citet{Lo2021ASignals}, reinforcing the interpretation of the selection effects as an overall normalisation constant to the Bayes factor, and highlighting the importance of the population model in interpreting the Bayes factor. 

However, in addition to validating previous derivations, we present our derivation for the posterior odds and the prior odds.
We argue that the posterior odds, which should be the ultimate arbiter for strong lensing detections, do not depend explicitly on selection effects for a fixed population model (Sec.~\ref{sec:odds_ratio}); this has implications for the detection of strong lensing. 
Furthermore, due to the form of the prior odds in the posterior odds, we argue that the Bayes factor, which depends on the observation time, needs to be interpreted with care and in the context of the prior odds.

\subsection{Evidence under the lensed hypothesis with selection effects} \label{ssec:evidence_lensed}

To answer if an event is strongly lensed or not, we need the definition of the strong lensing evidence, which enters into the Bayes factor. 
Here we derive the expression for the strong lensing evidence with selection effects, which we find to be consistent with the original derivation by~\citet{Lo2021ASignals}. 

Suppose a binary black hole is strongly lensed and thus split into $\Nimg$ images. 
The $\Nimg$ gravitational-wave images appear in the gravitational-wave detectors at times $t_i=t^{u}+t^{d}_i$. 
Since the separation in time can be quite large ($\mathcal{O}(\rm days - months)$) but the duration of the gravitational-wave detection is small ($\mathcal{O}(100 \, \rm ms)$), the strong lensing analysis is done on $\Nset=\Nimg$ separate sets of data $\data:=\{d_i\}$ around the event trigger times. 
Therefore, suppose that we have $\data$ 
sets of data that each include a gravitational wave. 
The strong lensing hypothesis asserts that the gravitational waves in this dataset arise from strongly lensed images of one binary black hole.

Presuming that we have a model for the detectable binary black holes and lenses in the Universe $p(\vtheta, \vthetal, \vy| \HL,\dett)$, we write the evidence under the lensing hypothesis as
\begin{equation} \label{eq:joint_evidence_conditioned}
\begin{split}
    p( \data |\dett,\HL) = \int &p(\data|\vtheta, \vthetal, \vy, \dett, \HL) \\
    &\times p(\vtheta, \vthetal, \vy|\dett, \HL) \dd \vtheta \dd \vthetal \dd \vy \,,
\end{split}
\end{equation}
where $p(\data|\vtheta, \vthetal, \vy, \dett, \HL)$ denotes the gravitational-wave likelihood where one only considers observable events, and $p(\vtheta, \vthetal, \vy|\dett, \HL)$ is the expected, observable population of black hole mergers and lenses in the Universe.  
Note that the population of \emph{detectable}, strongly lensed binary black hole mergers $p(\vtheta, \vthetal, \vy|\dett, \HL)$ is different from 
the \emph{underlying} population of strongly lensed binary black hole mergers $p(\vtheta, \vthetal, \vy|\HL)$. 

Note that both the gravitational-wave likelihood $p(\data|\vtheta, \vthetal, \vy, \dett, \HL)$ and the prior population $p(\vtheta, \vthetal, \vy|\dett, \HL)$ are conditioned on the detection in the joint evidence (Eq.~\ref{eq:joint_evidence_conditioned}). 
In particular, the strongly lensed binary black hole population conditioned on detection is 
\begin{equation}\label{eq:conditioned_prior}
    p(\vtheta, \vthetal, \vy|\dett, \HL) 
    = \frac{ P(\dett|\vtheta,\vthetal,\vy,\HL) p(\vtheta, \vthetal, \vy| \HL) }{ P(\dett|\HL) } \,,
\end{equation}
where $P(\dett|\vtheta,\vthetal,\vy,\HL)$ is the probability that a strongly lensed gravitational wave with $\Nimg$ images and parameters $\vtheta$, $\vthetal$, and $\vy$ produces $\Nimg$ separate, detectable gravitational-wave signals 
while 
\begin{equation}
    P(\dett|\HL) = \int P(\dett|\vtheta,\vthetal,\vy,\HL) p(\vtheta, \vthetal, \vy| \HL) \dd \vtheta \dd \vthetal \dd \vy\,,
\end{equation}
normalises the prior and is equivalent to the normalization  factor $\beta$ that enters the strongly lensed evidence in~\citet{Lo2021ASignals}. 
Note that the physical interpretation of the normalisation factor $P(\dett|\HL)$ is the prior probability of detecting a set of $\Nimg$ strongly lensed gravitational-wave signals. 

The argument in \citet{Lo2021ASignals} is that only the overall normalisation constant $P(\dett |\HL)$ enters the Bayes factor. 
In contrast, \citet{Liu2020IdentifyingVirgo} directly uses the population of \emph{detectable} strongly lensed black holes as a Malmquist before computing the evidence. 
That is, by evaluating the evidence (Eq.~\ref{eq:joint_evidence_conditioned}) with the strongly lensed binary black hole population model, conditioned on detection, $p(\vtheta, \vthetal, \vy|\dett, \HL)$, but without conditioning the likelihood (i.e., missing $P(\dett | \vtheta, \vthetal, \vy, \HL)$ in the denominator of the integrated in Eq.~\ref{eq:joint_evidence_conditioned}). 
Our results agree with \citet{Lo2021ASignals}. 
The approach adopted in \citet{Liu2020IdentifyingVirgo} would correspond to modifying the prior term in the evidence without modifying the likelihood. 
Although it would likely not significantly alter the conclusions of~\citet{Liu2020IdentifyingVirgo}, we believe the work may have missed modifying the likelihood function due to selection effects. 

In particular, the gravitational-wave likelihood conditioned with the detection of the $\Nimg$ binary black hole images is normalised differently than the likelihood without being conditioned on detection,
\begin{equation} \label{eq:conditioned_likelihood}
\begin{split}
    p(\data|\vtheta, \vthetal, \vy, \dett, \HL) &= \frac{p(\data|\vtheta, \vthetal, \vy, \HL)}{\int_{\data^\prime \in \data_\text{detectable}} p(\data^\prime|\vtheta, \vthetal, \vy, \HL) \dd \data^\prime}\,,\\
    &= \frac{p(\data|\vtheta, \vthetal, \vy, \HL)}{ P(\dett | \vtheta, \vthetal, \vy, \HL) }\,,
\end{split}
\end{equation}
that is, only the data space in which a detection would be made, $\data_\text{detectable}$, is considered. 

Inserting the conditioned likelihood (Eq.~\ref{eq:conditioned_likelihood}) and prior (Eq.~\ref{eq:conditioned_prior}) into the joint evidence, we find that the modification to the prior and the likelihood cancel out in such a way that only an overall normalisation constant remains 
\begin{subequations}
\begin{equation} \label{eq:joint_evidence_conditioned_1}
\begin{split}
	&p( \data |\dett,\HL) \\
	&= \int \frac{p(\data|\vtheta, \vthetal, \vy, \HL)}{ {P(\dett | \vtheta, \vthetal, \vy, \HL)} }\times  \frac{ {P(\dett|\vtheta,\vthetal,\vy,\HL)} }{P(\dett|\HL)} \\
    & \,\,\,\,\,\, \,\,\,\, \times p(\vtheta, \vthetal, \vy| \HL) \dd \vtheta \dd \vthetal \dd \vy \,, 
\end{split}
\end{equation}
\begin{equation} \label{eq:joint_evidence_conditioned_2}
\begin{split}
    &= \frac{1}{P(\dett|\HL)} \int p(\data|\vtheta, \vthetal, \vy, \HL)\\
    & \,\,\,\,\, \,\,\,\, \,\,\,\, \,\,\,\, \,\,\,\, \,\,\,\, \,\,\,\, \,\,\,\, \,\,\,\, \,\,\,\, 
    \times  p(\vtheta, \vthetal, \vy| \HL)  \dd \vtheta \dd \vthetal \dd \vy \,, 
\end{split}
\end{equation}
\begin{equation} \label{eq:joint_evidence_conditioned_3}
	\hspace*{-14em}
	= \frac{p(\data|H_L)}{P(\dett|\HL)} .\\
\end{equation}
\end{subequations}
After accounting for this missing factor in the likelihood, the evidence under the strong lensing hypothesis agrees with~\citet{Lo2021ASignals}. 
Indeed, the prior modification is cancelled out by the modification to the likelihood, such that only an overall normalisation constant remains; a similar line of argument is given in~\citep{Thrane2018AnModels} and related literature in a non-lensing context. 

Nevertheless, the correction may not be very significant, as the main normalisation constant $P(\dett|\HL)$ is already included in the Malmquist prior, such that integrand in the evidence (Eq.~\ref{eq:joint_evidence_conditioned_1}) is only missing a factor $P(\dett | \vtheta, \vthetal, \vy, \HL)$ in the denominator. 
However, as long as the detection is reasonably confident, the correction is expected to be small, because $P(\dett | \vtheta, \vthetal, \vy, \HL)\sim 1$ for most of the parameters in the posterior parameter space.\footnote{
	More specifically, the results could be reweighted with 
	$\langle P(\dett | \vtheta, \vthetal, \vy, \HL)^{-1}\rangle_{p(\vtheta, \vthetal, \vy|\data,\HL)}$, 
	but the correction would be order of unity as long as the detections are confident, i.e., 
	$P(\dett | \vtheta, \vthetal, \vy, \HL) = \int P(\dett|\vtheta, \vthetal, \vy, \vec n, \HL) \dd \vec n\sim 1$~\cite[see, e.g.][Eq.~E6]{Thrane2018AnModels}, where $\vec n$ denotes vector of noise realizations of the detectors. 
} 
Therefore, we believe it is unlikely that the correction could significantly alter the conclusions of~\citet{Liu2020IdentifyingVirgo}. 
Thus, while the work does include the first-order contribution from selection effects correctly, it is unlikely to be a major issue. 
However, the Bayes factor interpretation adopted by~\citet{Diego2021EvidenceLIGO-Virgo} does not include either of these selection effects, and we argue that it is therefore incorrect.

\subsection{Evidence under the unlensed hypothesis with selection effects}

One can perform a similar computation for the unlensed hypothesis $\HNL$, yielding
\begin{equation}\label{eq:conditioned_evidence_unlensed}
    p(\data|\dett, \HNL) = \frac{ p(\data|\HNL)}{P(\dett|\HNL)^{\Nset}} \,,
\end{equation}
where now $P(\dett|\HNL)$ is the prior probability of detecting an unlensed gravitational wave and is formally equivalent to the definition of $\alpha$ in~\citet{Lo2021ASignals}, while $\Nset(=\Nimg)$ is the number of datasets being analysed.

\subsection{The Bayes factor: Coherence ratio vs full treatment} \label{ssec:bayes_factor}

Dividing out the lensed (Eq.~\ref{eq:joint_evidence_conditioned_3}) and unlensed (Eq.~\ref{eq:conditioned_evidence_unlensed}) evidence yields the Bayes factor: 
\begin{equation} \label{eq:bludet_definition}
    \BLUdet = \frac{p(\data|\dett, \HL) }{p(\data|\dett, \HNL)} 
      = \frac{P(\dett| \HNL)^{\Nset} }{P(\dett| \HL)} \frac{p(\data| \HL) }{p(\data| \HNL)}\,,
\end{equation}
which has an overall normalisation constant and is consistent with the expression in~\citet{Lo2021ASignals}. 
It is worthwhile to point out that~\citet{Haris2018IdentifyingMergers, Hannuksela2019SearchEvents,Dai2020SearchO2} also used the Bayes factor without selection effects in their analysis: 
\begin{equation} \label{eq:blu_definition}
    \BLU = \frac{p(\data| \HL) }{p(\data| \HNL)}\,,
\end{equation}
where the overall normalisation constant is missing. 
However, the Bayes factor in these studies is interpreted as a ranking statistic that is then used in a frequentist-type p-value test, as opposed to the traditional interpretation of the Bayes factor as a detection criterion. 
In a p-value test, the overall normalisation of the ranking statistic cancels out, and therefore, the Bayes factor with and without selection effects would give the same result. 
Therefore, the conclusions in~\citet{Haris2018IdentifyingMergers,Hannuksela2019SearchEvents,Dai2020SearchO2} are unaffected by the discussion here.

However, even though selection effects enter the Bayes factor $\BLUdet$, the definition of the Bayes factor without selection effects $\BLU$ is quite relevant when considering strongly lensed detections. 
In particular, we argue that for a fixed population model, this latter definition of the Bayes factor $\BLU$ can be directly used in the posterior odds $\OLU$, which determines whether an event is ultimately lensed or not, and which does not depend on selection effects (Sec.~\ref{sec:odds_ratio}). 

\subsection{Cautionary note on selection effects}

Although the Bayes factor can be computed with or without selection effects, we should caution that the selection should be done carefully. 
In particular,~\cite{Cheung2023MitigatingMethods} and~\cite{Barsode:2024zwv} both find that in p-value type tests, the inclusion of the Malmquist prior changes the efficiency of frequentist-style strong lensing searches. 
In~\cite{Cheung2023MitigatingMethods}, the population prior $p(\theta)$ used in computing the Bayes factor was marginalised over a subset of the parameters $\vec \theta^\prime$ (in~\cite{Cheung2023MitigatingMethods} the $\vec \theta^\prime$ include detector-frame primary mass $m_1^z$ and mass ratio), and the Bayes factor was defined using marginalised priors.\footnote{This is equivalent to computing the Bayes factor assuming that the population prior factorises as 
$p(\theta) = p(\theta^\prime, \theta^\text{others}) \approx p(\theta^\prime)p(\theta^\text{others})$, where $\theta^\text{others}$ are the remaining parameters. } 
It would be tempting to assume that the marginalised priors are approximately the same 
on the basis that the full prior with or without selection effects is approximately the same in the confidently detected region of the parameter space.\footnote{
	I.e., assuming that $p(\theta^\prime) \approx \text{constant} \cdot p(\theta^\prime|\dett)$ because $p(\theta|\dett) = P(\dett|\theta) p(\theta)/P(\dett) \approx \text{constant} \cdot p(\theta)$ in the confidently detected region of the parameter space.
	}
However, even if the underlying population factorises as $p(\theta,\theta^\prime)=p(\theta)p(\theta^\prime)$, the detectable population does not factorise in general, i.e., $p(\theta,\theta^\prime|det) \neq p(\theta|\dett) p(\theta^\prime | \dett)$.\footnote{
	An example of this is the detectable population of primary component masses $p(m_1|\dett)=\int p(\theta|\dett)p(\theta^\text{others}) \dd \theta^\text{others}$, with $\theta^\text{others}$ marking all other parameters, which has a strong bias towards towards higher masses, while $p(m_1)=\int p(\theta) p(\theta^\text{others}) \dd \theta^\text{others}$ does not.} 
Marginalising over any subset of the parameters in the Bayes factor, likewise, would result in biases. 
Therefore, the efficiency loss found in~\cite{Cheung2023MitigatingMethods} from using the intrinsic population is likely due to this reason. 
We suspect a similar reason explains the results in~\cite{Barsode:2024zwv}. 

One way to correct issues related to prior marginalisation would be to include the full correlations in the prior when reconstructing the population, for example, by using the non-parametric models advocated in~\cite{Cheung2023MitigatingMethods}. 
Another method would be to abandon the Bayes factor approach and instead use the Bayes factor as a ranking statistic in a frequentist-style test, as in~\cite{Barsode:2024zwv}. 
However, how the population reconstruction should be done in a way that introduces no biases is still an open question. 

\section{Prior odds} \label{sec:prior_odds}

The prior odds $P(\HL|\dett)/P(\HNL|\dett)$ is another quantity shrouded in some mystery. 
In particular, in several prior works, including some of the authors' works, the prior odds was taken to be the probability that a given individual event is strongly lensed versus that it is not~\citep{Haris2018IdentifyingMergers,Hannuksela2019SearchEvents, Liu2020IdentifyingVirgo}. 
However, the prior odds should instead be defined as the ratio of the probability that a given \emph{set} of events are all strongly lensed images of the same source to the probability of the converse, rather than the ratio of the probability that a given individual event is lensed to the probability that it is not.\footnote{
	In fact, this realisation was made already before in the~\cite{TheLIGOScientificCollaboration2021SearchRun}, which originally contained a dedicated appendix in the paper to this end, but there were still ambiguities in the derivation at the time. } 
This distinction significantly changes the interpretation of the prior odds. 

\subsection{Intuitive argument why prior odds $\neq$ relative rate}

The prior probability of lensing is loosely related to the well-known birthday problem. 
In probability theory, the birthday problem asks for the probability that in a set of $n$ randomly chosen people, at least two will share the same birthday. 
The birthday paradox is the counterintuitive fact that only 23 people are needed for that probability to exceed 50\%. 
The underlying reason can be conceptualized by noting that there are $23\times 22/2=253$ \emph{pairs} to consider, which makes the probability that two people have the same birthday much higher than the probability that two people have the same birthday, multiplied by the number of people. 
In the case of strong gravitational-wave lensing, there are likewise $N$ gravitational-wave signals. Still, there are significantly more pairs (or more generally, sets) of signals that we need to analyse. 
Even if the chance that two signals appear similar within detector accuracy is small, the probability increases rapidly as our catalogue of individual signals grows. 

With this context, let us expand upon some of the previous arguments presented for the prior odds, starting from the relative rate of lensing. 
The relative rate of lensing in the current ground-based gravitational-wave detectors at various sensitivity levels using either simulated or estimated population of lenses and binary black holes has been quantified in several works~\citep{Ng2017PreciseHoles, Li2018GravitationalPerspective, Oguri2018EffectMergers, Buscicchio2020ConstrainingBackground, Mukherjee2020InferringBackground, TheLIGOScientificCollaboration2021SearchRun, Xu2021PleasePopulations, Wierda2021BeyondLensing,Wempe2022AObservations}, typically arriving at relative rates of detectable events of around $\sim 1:1000$.\footnote{Some variability in the rate estimates is due to different assumptions for the lens population and especially the merger-rate density of binary black holes at high redshift.} 
Then, neglecting the number of images produced by the lenses, if we detect $\Ndet$ gravitational-wave events, on average, about $\sim 1/1000$ are lensed. 

Thus, one could argue that any given random event that is strongly lensed is about one in a thousand, or equal to the relative rate of detected lensed events: 
\begin{equation} \label{eq:incorrect_prior_odds}
\frac{P(\text{an individual event is lensed}|\dett)}{P(\text{an individual event is not lensed}|\dett)} \sim \frac{\Rldet}{\Rudet} \,.
\end{equation}
where $\Rldet$ is the rate of detectable lensed events per year and $\Rudet$ is the rate of detectable unlensed events per year. 
Indeed, this argument has motivated many previous works to adopt the relative rate of lensing as the prior odds.
However, we argue that this is incorrect, because it neglects the fact that every Bayesian analysis of strong lensing analyses \emph{sets} of signals, instead of individual signals. 
Indeed, as we will see, formulating the prior probability using a \emph{set} of signals will produce a significantly lower prior probability of lensing, giving the same counterintuitive result as the birthday problem: The probability of matching two pairs by accident is significantly greater than the probability of matching individual pairs of events multiplied by the number of events. 

\subsection{Formal derivation of prior odds} \label{ssec:prior_odds}

The challenge in considering the prior odds as the probability that an individual event is lensed is that the strong lensing analysis analyses \emph{set} of events, not individual events. 
Here we find the counterintuitive result that the prior probability of lensing decreases as the number of gravitational-wave events that we analyse increases, which is the Bayesian version of the frequentist finding of rapidly rising false alarms in~\citep{Calskan2022LensingWaves}. 

\subsubsection{Number of lensed and unlensed gravitational waves}

Let us suppose that there are $\N$ gravitational-wave signals in the detector (including signals that are not detectable due to residing below the noise level of the detector). 
The total number of signals is the sum of lensed and unlensed signals
\begin{equation}
    \N=\Nu + \Nimg \Nl\,,
\end{equation}
where, for simplicity, we assume a fixed number of images $\Nimg$. 
The $\Nu$ is the number of unlensed signals, while $\Nl$ is the number of \emph{sets} of lensed signals. 
Note that $\Nl$ does not represent the total number of strongly lensed signals. 
The total number of mergers that are strongly lensed, $\Nl$, create:
\begin{equation} \label{eq:number_lensed_signals_not_mergers}
    N_L^\text{signals}=\Nl \Nimg \,
\end{equation}
strongly lensed signals, corresponding to $N_L$ lensed mergers.

The number of lensed and unlensed events follows a Poissonian distribution
\begin{equation}
\begin{split}
    P(\Nl) &= \frac{\Nlavg^{\Nl} e^{-\Nlavg}}{\Nl!}\,, \\
    P(\Nu) &= \frac{\Nuavg^{\Nu} e^{-\Nuavg}}{\Nu!}\,, \\
\end{split}
\end{equation}
where 
\begin{subequations}\label{eq:nl_nu_avg_definition}
    \begin{equation} \label{eq:rl_definition}
        \Nlavg=\Rl T
    \end{equation}
    \begin{equation} \label{eq:ru_definition}
        \Nuavg=\Ru T
    \end{equation}
\end{subequations}
are the expected number of lensed and unlensed events within observation time $T$, and $\Rl$ and $\Ru$ are the expected rates of lensed and unlensed events. 
For the sake of simplicity, we assume that these rates are given as prior knowledge, although technically one could also allow them to vary. 

\subsubsection{Prior probability of lensing and prior odds} \label{sssec:prior_odds}
Since there are $\Nl$ sets that are strongly lensed counterparts to each other and $\binom{\N}{\Nimg}$ total sets of signals, the probability that an event is a strongly lensed set, given a fixed image number and number of lensed/unlensed events, 
\begin{equation}
    P(\HL|\Nl, \Nu) =\frac{\Nl}{\binom{\N}{\Nimg}} \,.
\end{equation}
When the average number of lensed sets of events, $\Nlavg$, and unlensed events $\Nuavg$, are known, we can approximate
\begin{equation}
    P(\HL|\Nlavg, \Nuavg) =\frac{\Nlavg}{\binom{\Navg}{\Nimg}} \,,
\end{equation}
where $\Navg=\Nuavg+\Nimg \Nlavg$. 
We can expand this as\footnote{Note that $P(\HL|\Nlavg, \Nuavg)$ can be alternatively written in terms of the expected rate of lensed and unlensed events, $\Rl$ and $\Ru$, respectively, and the observation time $T$, i.e., $P(\HL|\Nlavg, \Nuavg)=P(\HL|\Rl,\Ru,T)$.}
\begin{equation}
    P(\HL|\Nlavg, \Nuavg) \approx \Nimg! \frac{\Nlavg}{\Navg^{\Nimg}} \left[ 1 + \mathcal{O}\left (\Navg^{-1}\right) \right] \,.
\end{equation}
In the limit of a high number of mergers, and assuming that the number of images is much smaller than the total number of signals,  
the prior probability of lensing reduces to
\begin{equation}
    P(\HL|\Nlavg, \Nuavg) \approx \Nimg! \frac{\Nlavg}{\Navg^{\Nimg}} \,.
\end{equation}
Note that the prior probability of lensing depends sensitively on the number of events $\N$. 
The more events we observe, the lower the prior probability of strong lensing.

With the definition of the prior probability of lensing $P(\HL)$, and assuming that the probability that a given set is strongly lensed is much smaller than the probability that it is not (i.e., $P(H_U)\approx 1$), the prior odds without selection effects 
\begin{equation}
\begin{split}\label{eq:priorodds}
    &\PLU=\frac{P(\HL)}{P(\HNL)} \\
    &\approx \Nimg! \frac{\Nlavg}{\Navg^{\Nimg}} \left[ 1 + \mathcal{O}\left (\Navg^{-1}\right) \right]  \\
    &\approx \Nimg! \frac{\Nlavg}{\Navg^{\Nimg}}  \,,
\end{split}
\end{equation}
where, in the last line, we took the limit of a high number of mergers. 
Note that the prior odds reduce to the relative rate of lensing only in single-image analyses ($\Nimg=1$).

\begin{figure}
    \centering
    \includegraphics[width=\columnwidth]{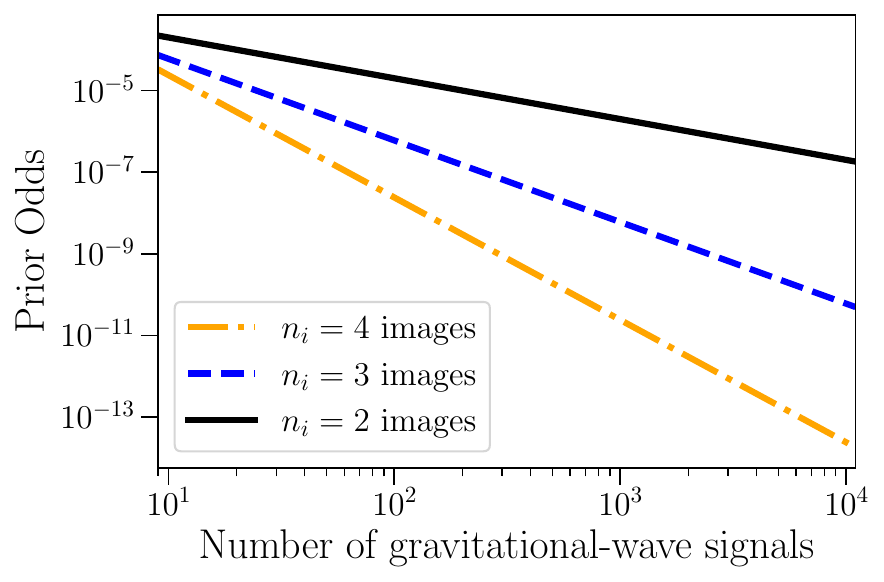}
    \caption{The prior odds $P^L_U$ in favour of lensing as a function of the number of gravitational-wave signals $N$ for a variable number of strong lensing image configurations. 
    Particularly noteworthy is that the prior odds in favour of lensing \emph{decrease} as the number of events increases (or as time passes). 
    Indeed, the prior probability that any given set of data contains a strongly lensed signal decreases over time. 
    However, the arrival time priors (arrival time odds) increase so that they compensate for this decrease, leaving the posterior odds intact and independent of the number of events.
    The ratio of lensed-to-unlensed events is fixed at one in a thousand.
}
    \label{fig:prior_odds}
\end{figure}

The curious result is that the prior probability in favour of lensing \emph{decreases} over time, as the number of events increases (Fig.~\ref{fig:prior_odds}).\footnote{In particular, note that $\Navg=(\Ru+\Nimg \Rl)T$.}
Furthermore, the form of the prior odds suggests that the probability that we can detect any strong lensing becomes extremely small once we reach a certain number of events, which, at first glance, is catastrophic to confident detections. 
For example, for an observing time $T=1 \, \rm yr$, rates $R=1000 \, \rm yr^{-1}$ and $\Rl=1 \, \rm yr^{-1}$, and several images $\Nimg =4$, the probability that any given set of $\Nimg$ events is lensed, before we have analysed the data, $\PLU \approx 2 \times 10^{-11}$, which is extremely small. 

Indeed, the decreasing probability of the lensed hypothesis is consistent with the findings of~\citet{Calskan2022LensingWaves}, which demonstrates that the probability of detecting lensing becomes less and less likely as time passes, or as the number of events we observe increases, due to false alarms or \textquotedblleft lensing mimickers.\textquotedblright
However, the time-delay prior used in the lensing hypothesis could potentially balance out the rising probability of false alarms, noted in~\citep{Wierda2021BeyondLensing}. 
Indeed, we will derive this finding using the Bayesian formalism in the next section.

\subsection{Selection effects}

To incorporate selection effects, one must account for the fact that only a fraction of events will be detected. 
In particular, one  must replace the expected number of lensed and unlensed mergers with the expected number of detectable lensed and unlensed mergers
\begin{equation*}
    \begin{split}
        \Nlavg &\rightarrow \Nlavg P(\dett|\HL)\,, \\
        \Nuavg &\rightarrow \Nuavg P(\dett|\HNL)\,.
    \end{split}
\end{equation*}
Since the unlensed mergers dominate the total number of mergers, the prior odds with selection effects 
\begin{equation}
\begin{split}\label{eq:prioroddsdet}
    &\PLUdet=\frac{P(\HL|\dett)}{P(\HNL|\dett)} = \frac{P(\dett|\HL)}{P(\dett|\HU)^{\Nimg}} \PLU \,.
\end{split}
\end{equation}
Thus, the selection effects introduce an overall normalisation constant. 

In layman's terms, the prior probability of strong lensing, conditioned on the detection, $\PLUdet$, is approximately the relative probability that any given observed set of signals is strongly lensed. 
In the meantime, the prior probability of strong lensing without this selection is the corresponding probability for the intrinsic population of strongly lensed gravitational waves. 
However, as we will see, the terms related to the selection effects cancel out in the posterior odds, so it will, in principle, not matter whether or not the selection effects are included in the posterior odds or not, as long as it is done consistently in both the Bayes factor and the prior odds.

\section{Posterior odds} \label{sec:odds_ratio}

The strong lensing posterior odds quantifies how likely any given event is strongly lensed. 
It includes the Bayes factor and the prior belief in the lensed and unlensed hypotheses. 
Here we find the following: 
(i) selection effects cancel out in the posterior odds for a fixed population model, and 
(ii) the decrease in the prior odds $\PLUdet$ is compensated for by the increase in the time-delay prior ratio, in the limit of a high number of mergers.

\subsection{Selection effects cancel out in the posterior odds}

Let us start with the definition of the posterior odds
\begin{equation}
    O^L_{U} = \frac{p(\data | \HL, \dett )}{p(\data | \HNL, \dett )} \frac{P(\HL|\dett)}{P(\HNL|\dett)}\,,
\end{equation}
where the first term, the Bayes factor, can be decomposed as 
\begin{equation}
\begin{split}
    \mathcal B^L_{U} &=  \frac{p(\data | \HL, \dett )}{p(\data | \HNL, \dett )} \\
    &=  \frac{P(\dett | \HNL )^{\Nimg}}{P(\dett | \HL )} \frac{p(\data | \HL)}{p(\data | \HNL)}\,,
\end{split}
\end{equation}
and the second term, the prior odds, can be decomposed as 
\begin{equation}
\begin{split}
    \frac{P(\HL|\dett)}{P(\HNL|\dett)} = \frac{P(\dett | \HL )}{P(\dett | \HNL )^{\Nimg}} \frac{P(\HL)}{P(\HNL)}\,.
\end{split}
\end{equation}
Combining the prior odds with the Bayes factor, we find a rather intriguing result, namely that the posterior odds does not depend on the selection effects:
\begin{equation}
\begin{split}
    O^L_{U} &= \BLUdet \PLUdet \\
    &= \left( \frac{{P(\dett | \HNL )^{\Nimg}}}{{P(\dett | \HL )}}\frac{p(\data | \HL)}{p(\data | \HNL)} \right) 
    \left(  \frac{{P(\dett | \HL )}}{{P(\dett | \HNL )^{\Nimg}}} \frac{P(\HL)}{P(\HNL)} \right) \\
    &= \frac{p(\data | \HL)}{p(\data | \HNL)} \frac{P(\HL)}{P(\HNL)} \\
    &= \BLU \PLU \,,
\end{split}
\end{equation}
that is, the selection effects in the prior odds cancel out with the selection effects in the Bayes factor for a fixed population model. 
Nevertheless, it is still entirely valid to include the selection effects in the Bayes factor computation, as long as they are also included in the computation of the prior odds.
Thus, our work does not argue against the earlier approaches, which included the selection effects in the Bayes factor, but merely highlights the importance of including the prior odds in detection claims. 

In the expression without selection effects, the rate of lensed and unlensed events is the rate of \emph{intrinsic} mergers, instead of the rate of detectable mergers. 
This simplifies the computation of the prior odds significantly. 
Indeed, in terms of strong lensing statistical analyses such as~\citet{Ng2017PreciseHoles, Li2018GravitationalPerspective, Oguri2018EffectMergers, Buscicchio2020ConstrainingBackground, Mukherjee2020InferringBackground, TheLIGOScientificCollaboration2021SearchRun, Xu2021PleasePopulations, Wierda2021BeyondLensing,Wempe2022AObservations, More2021ImprovedEvents,Smith2022DiscoveringObservatory, Phurailatpam2024LerSimulator}, determining the intrinsic rate of mergers is significantly less computationally expensive than determining the rate of detectable mergers. 

However, a challenge still needs to be addressed: The prior odds keep decreasing as we observe more events. 
We address this in the next subsection, where we show that the decrease in the prior odds over time is compensated by the increase in the Bayes factor (particularly in the ratio of the arrival time priors) over time. 

\subsection{The increase in the time-delay prior ratio cancels the decrease in the prior odds as the number of gravitational-wave events increases}
\noindent
In the prior odds section (Sec.~\ref{sec:prior_odds}), we found the result that the prior odds in favour of strong lensing decreases as the number of events we detect increases 
\begin{equation}
    \PLU \approx \Nimg! \frac{\Nlavg}{\Navg^{\Nimg}}  \left[ 1 + \mathcal{O}\left (\Navg^{-1}\right) \right] \,.
\end{equation}
At first glance, the prior odds would seem to make the detection of strong lensing impossible once the number of events has reached a certain value. 
This is because the prior odds keep decreasing as the number of gravitational-wave observations increases. 
However, as we will see, this is not a problem for detection, as the time-delay priors in the Bayes factor compensate for this decrease in the prior odds.

In particular, separating the Bayes factor into two components, similar to the approach in~\citet{Haris2018IdentifyingMergers} (see also~\citep{More2021ImprovedEvents}), we have 
\begin{equation}
    \BLU = \widetilde \BLU  \RLU \,,
\end{equation}
where, the ratio of the arrival time priors evaluated at the trigger time (hereafter \textquotedblleft arrival time odds\textquotedblright), 
\begin{equation} \label{eq:rlu_defnition}
    R^L_{U} = \left(\frac{p(\vec{t}|\HL)}{p(\vec{t}|\HNL)} \right)_{\vec{t}=\vec{t}_{\rm gps}}\,.
\end{equation}
The arrival time priors are taken out from the Bayes factor $\BLU$ to obtain $\widetilde \BLU$. 
This can be done because the lensed and unlensed time-delay prior is not sensitive to fluctuations at these scales $p(\vec{t}+\vec{\delta t}|\HL/\HNL)\approx p(\vec{t}|\HL/\HNL)$, where $\delta t \sim$ millisecond is the accuracy of the detectors in determining the arrival time of the signal.

Given an observation time $T$, the unlensed arrival time prior follows a time-ordered Poissonian distribution~\citep{Lo2021ASignals}
\begin{equation}
    p(\vec{t}|\HU) = \Nimg! T^{-\Nimg} \,.
\end{equation}
Within a given observation time $T$, there are on average 
\begin{equation} \label{eq:Rtot_definition}
    \Navg=\Rtot \times T\,,
\end{equation}
gravitational-wave events, where $R=\Rl + \Ru$ is the total rate of mergers. 
Therefore,  
\begin{equation}
    p(\vec{t}|\HU) = \Nimg! R^{\Nimg} \Navg^{-\Nimg}  \,.
\end{equation}

On the other hand, the lensed time-delay arrival time prior is a Poissonian distribution for the first event. The subsequent time-delay prior follows predictions from lensing statistics. 
Thus, the strongly lensed arrival-time prior 
\begin{equation}
\begin{split}
    p(\vec{t}|\HL) &= p(t_0|\HL) p(\Delta \vec{t}|\HL)
\end{split}
\end{equation}
where $t_0=t_u+t_0^d$ is the arrival time of the first gravitational-wave image and $\Delta \vec{t}:=\{t_1^d-t_0^d,t_2^d-t_1^d,t_3^d-t_2^d\}$ is the arrival time difference between the subsequent images.\footnote{We define the lensing time-delay prior such that the images are time-ordered ($t_i>t_{i+1}$).} 
Although the arrival time difference $p(\Delta \vec{t}|\HL)$ depends on the lens and the source model~\citep{More2021ImprovedEvents}, it is independent of the total number of events~\citep[e.g.][]{Wierda2021BeyondLensing}. 
The arrival time of the first gravitational-wave signal $p(t_0|\HL)$ again follows a Poissonian distribution. 
Therefore, 
\begin{equation}
    p(\vec{t}|\HL) = \frac{1}{T} p(\Delta \vec{t}|\HL) \,.
\end{equation}
Similarly to the unlensed case, we can express the arrival time prior in terms of the expected, intrinsic rate $R$ and the expected number of events $\Navg$ 
\begin{equation}
    p(\vec{t}|\HL) = R \Navg^{-1} p(\Delta \vec{t}|\HL) \,.
\end{equation}
Combining the lensed and unlensed time-delay priors, the ratio of the time-delay priors
\begin{equation}
    \RLU  = \Navg^{\Nimg-1} \frac{p(\Delta \vec{t}|\HL)}{\Nimg! R^{\Nimg-1}}  \,,
\end{equation}
Intriguingly, here we observe the opposite of what we observed in the prior odds: 
The ratio of the arrival time priors (arrival time odds) \emph{increases} as we observe more events. 

Indeed, combining the ratio of the arrival time priors with the prior odds in the final posterior odds, we find that the increase in the arrival time odds exactly compensates for the decrease in the prior odds. 
In particular, in the limit of a large number of events, the posterior odds
\begin{equation} \label{eq:posterior_odds_split}
\begin{split}
    O^L_{U} &\approx  \widetilde \BLU \times \TLU \,,
\end{split}
\end{equation}
where\footnote{Note that the below $\TLU$ is dimensionless since $[p(\Delta \vec t|\HL)]=[\Delta t]^{-\Nimg+1}$}  
\begin{equation}\label{eq:tlu_definition}
   \TLU \approx  \frac{\Rl }{R}  \frac{p(\Delta \vec{t}|\HL)}{R^{\Nimg-1}} \,,
\end{equation}
encodes the information of both the prior odds and the arrival time odds (hereafter we call $T^L_U$ rate odds) 
and 
\begin{equation} \label{eq:blu_tilde_definition}
\begin{split}
   \widetilde \BLU &= \BLU/\RLU \\
   &= \frac{p(\data|\HL)}{p(\data|\HU)} \left( \frac{p(\vec{t}|\HU)}{p(\vec{t}|\HL)} \right)_{\vec t=\vec t_\text{gps}}\,,  \\
\end{split}
\end{equation}
is the Bayes factor with the arrival time priors and selection effects divided out, such that it no longer depends on the observation time $T$.
\emph{Note that the expression is equivalent to the Bayes factor evaluated using a uniform arrival time prior. }

\subsection{Example $\TLU$ calculation} \label{sec:validation}

\begin{figure*}
    \centering
    \includegraphics[width=\textwidth]{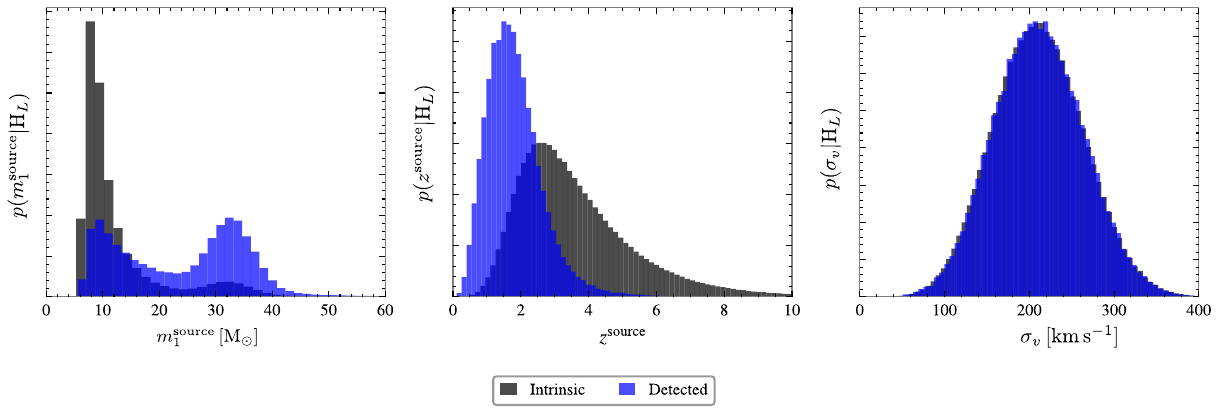}
    \caption{A summary of our strongly lensed population model's mass, redshift, and velocity dispersion priors. 
    The mass model follows the \textsc{Powerlaw+Peak} model with a primary mass slope index $\alpha=3.63$ and mass ratio power-law index $\beta_q=1.26$ (left panel), the redshift distribution traces the merger-rate density convolved with the strong lensing optical depth, the velocity dispersion of the galaxies follows the SDSS galaxy catalog (right panel) and we assume a simple singular isothermal sphere (SIS) lens model for illustrative purposes. 
    The population model and the simulation procedure follow the \texttt{ler} gravitational-wave lensing package~\citet{Phurailatpam2024LerSimulator}. 
    These priors are used to evaluate the ratio of arrival time priors in the lensed-to-unlensed hypothesis and the rate of strong lensing detections. 
    }
    \label{fig:population_model_summary}
\end{figure*}

\begin{figure*}
    \centering
    \includegraphics[width=0.6666\textwidth]{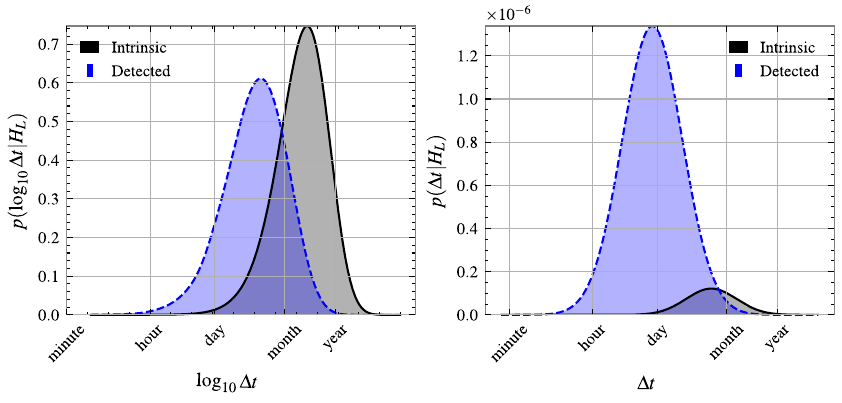}
    \caption{Distribution of the expected lensing time-delay differences induced by a population of strong lenses and a binary black hole population tracing the star-formation rate density. 
    Here, we presume a simple singular isothermal sphere (SIS) lens model, which creates two strong lensing images, as an easy illustrative example. 
    We will use the time-delay distribution to estimate the posterior odds. 
    }
    \label{fig:time_delay_sis}
\end{figure*}

Here is an example use case of the strong lensing posterior odds in a simple scenario. 
In particular, we simulate strongly lensed gravitational-wave events produced by a population of stellar remnants tracing the star-formation rate density as predicted by population synthesis codes~\citep[see][]{Oguri2018EffectMergers}, following the \textsc{Powerlaw+Peak} mass model~\cite{TheLIGOScientificCollaboration2021GWTC-3:Run}, and lensed by a population of singular isothermal sphere (SIS) lenses with parameters approximately following the expectations of the SDSS galaxy catalog~\cite{SDSS} (Fig~\ref{fig:population_model_summary}). 
The unlensed population model is the same, except that the lens population is absent. 
The population model and the simulation procedure follow~\citet{Phurailatpam2024LerSimulator}, except we use the SIS lens model for simplicity. 
Therefore, we refer the interested reader to~\citet{Phurailatpam2024LerSimulator}. 
This model expects the intrinsic strong lensing time delays at $\sim \rm hours - months$ apart (Fig.~\ref{fig:time_delay_sis}). 

\begin{figure}
    \centering
    \includegraphics[width=\columnwidth]{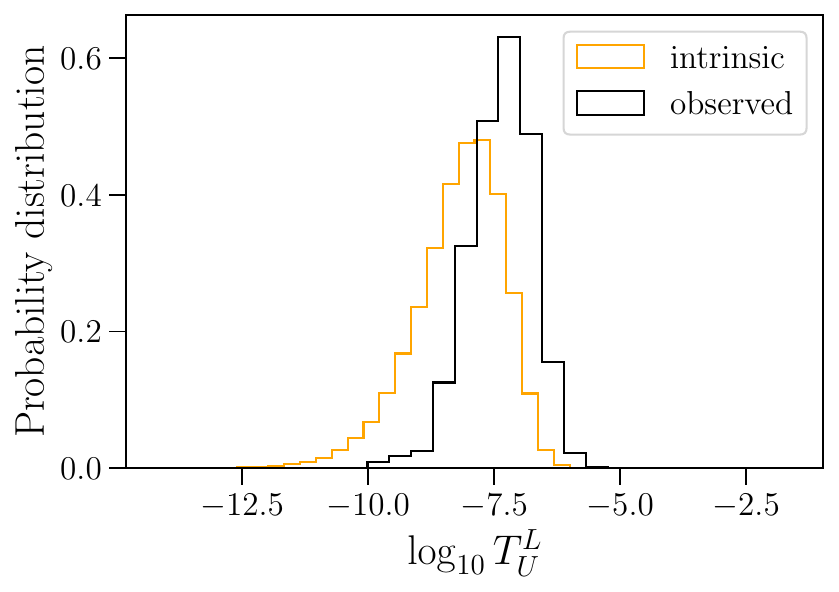}
    \caption{Distribution of the expected distribution of $T^L_U$ for both observed (black) and intrinsic (orange) population of strongly lensed binary black holes formed from stellar merger remnants. 
    The singular isothermal sphere model used here creates double images. 
    For us to classify an event as a detection, the time-independent Bayes factor without selection effects $\widetilde{B}^L_U$ must overcome the prior odds against the rate odds $T^L_U$. 
    Indeed, a large, strong lensing Bayes factor is insufficient as a detection criterion. 
    }
    \label{fig:tlu_distribution_sis}
\end{figure}

Fig.~\ref{fig:tlu_distribution_sis} shows the expected distribution of the rate odds $T^L_U$, which encapsulates the combined effects of the prior odds and the arrival time priors. 
The distribution is shown for the observed (black) and intrinsic (orange) populations of strongly lensed binary black holes formed from stellar merger remnants, assuming a singular isothermal sphere lens model that produces double images. 
This distribution is crucial for interpreting the posterior odds, as it sets the threshold for detection. 
Specifically, for a strong lensing detection to be confidently claimed, the time-independent Bayes factor without selection effects $\widetilde{B}^L_U$ must exceed the rate odds $T^L_U$. 
This underscores that a large Bayes factor alone is insufficient; the rate odds must also be favourable, reflecting the interplay between the population model, lensing time-delay priors, and the intrinsic rates of lensed and unlensed events.

\subsection{Implications for detections}

The overarching conclusion is that the arrival time odds $\RLU$ increases over time, while the prior odds $\PLU$ decreases over time. 
The two quantities compensate for each other in a precise manner. 
Therefore, the posterior odds are independent of the number of events and depend only on the intrinsic rate of lensed and unlensed mergers. 
Thus, any detection claim based on Bayesian analysis should evaluate the posterior odds $O^L_U=\widetilde{B}^L_U T^L_U$, which consist of a time-independent Bayes factor (without the arrival time priors) $\widetilde{B}^L_U$ and the rate odds $T^L_U$.
An attempt to quantify a detection using the time-dependent Bayes factor $B^L_U$ without incorporating the prior odds would result in an overtly optimistic estimate for detection.

It is furthermore important to stress that the lensing time-delay prior $p(\Delta \vec t|\HL)$ regularises the number of pairs that ever need to be considered. 
This is because there is a maximum $\Delta \vec t$ predicted by strong lensing, such that $p(\Delta \vec t|\HL)$ has a finite domain. 
For example, if we have 10 years of CE data, we need only consider $\sim 10\times (R\cdot 1 \, \text{yr})^2$ pairs as opposed to $(R\times 10 \, \text{yr})^2$ pairs. 
This finding is consistent with~\cite{Wierda2021BeyondLensing}. 

There is furthermore a common belief that the rising number of gravitational-wave detections would make detections less and less likely. 
However, this is not the case, as the posterior odds is independent of the number of events. 

\section{Discussion and conclusions} \label{sec:discussion}

Our analysis clarifies several key aspects of Bayesian strong lensing detection in gravitational-wave astronomy, resolving discrepancies in earlier interpretations and emphasising the role of population models and selection effects. Below, we summarise our findings and their implications.

Our key results are the following:
\begin{itemize}
    \item \emph{Prior odds and observation time}: The prior odds decrease as more gravitational-wave events are detected, reflecting the growing improbability that any specific set of events is lensed. However, this does not preclude detection, as the Bayes factor compensates for this effect.
    \item \emph{Arrival time priors and Bayes factor}: The arrival time prior ratio (and thus the Bayes factor) increases with observation time. This counterbalances the decreasing prior odds, ensuring that the posterior odds remain stable.
    \item \emph{Posterior odds as the robust statistic}: 
    The posterior odds are independent of observation time or catalogue size, making it the definitive criterion for strong lensing detection. 
    \item \emph{Selection effects}: For a fixed population model, the posterior odds can be computed with or without explicit selection effects in its factors, as they cancel out, as long as it is done consistently. 
    \item \emph{The necessity of binary black hole population models} The strong lensing hypothesis inherently tests whether two events are more likely to be lensed counterparts than random coincidences. Failure to account for the population model will result in an overtly optimistic or pessimistic Bayesian estimate of detection; our conclusions agree with~\cite{Lo2021ASignals}.
    \item \emph{Inclusion of lensing time delays}: The strong lensing time-delay model is critical to detection. If the strong lens time delay priors are not included, the ratio of arrival time priors does not cancel the prior odds in the posterior odds, making detections impossible. That is, the posterior odds $[\OLU]_\text{no lensing time delay} = [\BLU \PLU]_\text{no lensng time delay}\rightarrow 0$ since $\PLU\rightarrow 0$ in the limit of a large number of detections while $\BLU$ is constant, if the arrival time odds do not include a statistical model for the lens population, in agreement with~\citet{Calskan2022LensingWaves,Wierda2021BeyondLensing}.
    \item \emph{Possibility of strong lensing detections}: Even though the number of gravitational-wave detections is growing, the posterior odds is independent of it. Thus, the number of gravitational-wave events in our catalogue does not impact strong lensing detection claims. 
\end{itemize}

In terms of previous works, our results are largely in agreement with~\citep{Lo2021ASignals}; we confirm that, if the posterior odds are computed without selection effects, both the ratio of arrival time priors and the prior odds should be computed for the \emph{intrinsic} population of strong lenses, not for the observable population of strong lenses. We find that our results disagree slightly with~\citep{Liu2020IdentifyingVirgo}, which includes the selection effects by replacing the intrinsic prior with a Malmquist prior, but without modifying the likelihood. 
However, in addition to these works, we derive the posterior odds, which should ultimately be used to quantify detection
Our results disagree with \citet{Diego2021EvidenceLIGO-Virgo}, which advocate for a \textquotedblleft coherence ratio\textquotedblright (Bayes factor without selection effects or population priors) as the primary lensing statistic. 
While such a ratio may seem agnostic, it ignores two critical factors: (i) Selection Effects: Omitting selection effects mislabels the Bayes factor unless compensated in the prior odds. 
(ii) Time-delay model: Without a lens time-delay model, the arrival time prior cannot counteract the prior odds, rendering detection statistically impossible for large catalogues due to the birthday problem. 
The use of a time-delay prior is necessary, but some care is required to ensure that the prior contains all available information.

In terms of quantifying the detectability of these strongly lensed gravitational waves, there exist recent, seminal frequentist approaches~\citep{Barsode:2024zwv}, which employ similar strategies to~\cite{Haris2018IdentifyingMergers}. 
In these works, the trials factor originating from the many pairs of events is included. 
Because the Bayes factor is used as a ranking statistic against a background, we would not expect the \textquotedblleft birthday problem\textquotedblright nor the discussion on selection effects to alter the robustness of detection claims using these methods.\footnote{It is however worth noting that the selection effects can change how optimal the ranking statistic being used in these frequentist tests is. } 
Furthermore, since detection is quantified using p-value type tests (by treating the Bayes factor as a ranking statistic), an incorrect population model in the Bayes factor definition does not immediately lead to a false claim of lensing detections, although an incorrect population model in constructing the background for the ranking statistic can. 
The background requires only knowledge of the unlensed population, and so it is robust to uncertainties in the lensed population -- though an incorrect population model in the ranking statistic would lead to sub-optimal detection efficiency. 
For realistic detection estimates, using frequentist-style methods, we therefore refer the interested reader to~\cite{Barsode:2024zwv}.

\section*{Acknowledgements}

O.A.H. and H.P. acknowledge support by grants from the Research Grants Council of Hong Kong (Project No. CUHK 14304622, 14307923, and 14307724), the start-up grant from the Chinese University of Hong Kong, and the Direct Grant for Research from the Research Committee of The Chinese University of Hong Kong. 
H.N. and C.V.D.B are supported by the research program of the Netherlands Organisation for Scientific Research (NWO)
J.D.E.C. acknowledge the support of NSF Grant PHY-2513124. 
We thank Ilya Mandel for insightful email exchange on selection effects. 
We thank Ankur Barsode and Srashti Goyal for enlightening discussions relating to the frequentist-type methods to detect lensing as well as on these Bayesian derivations, and Ankur for pnp-reviewing the draft. 
We furthermore thank Rico Ka-Lok Lo for a number of comments that helped to significantly improve the work.


\end{document}